\documentclass[aps,superscriptaddress,showpacs,prd,twocolumn]{revtex4}
\usepackage{hyperref}
\usepackage{graphicx}
\usepackage{amssymb}
\usepackage{amsmath}
\usepackage{epstopdf}
\usepackage{color}
\usepackage{mathrsfs}
\usepackage{braket}
\begin{document}

 \title{Vanishing Geometric Discord in Non-Inertial Frames}
\author{Eric G. Brown}
 \affiliation{Department of Physics and Astronomy, University of Waterloo, Waterloo, Ontario N2L 3G1, Canada}
 \author{Kyle Cormier}
 \affiliation{Department of Physics and Astronomy, University of Waterloo, Waterloo, Ontario N2L 3G1, Canada}
 \author{Eduardo Mart\'{i}n-Mart\'{i}nez}
   \affiliation{Department of Physics and Astronomy, University of Waterloo, Waterloo, Ontario N2L 3G1, Canada}
  \affiliation{Institute for Quantum Computing and Department of Applied Mathematics, University of Waterloo, 200 University
Avenue W, Waterloo, Ontario, N2L 3G1, Canada}
 \author{Robert B. Mann}
  \affiliation{Department of Physics and Astronomy, University of Waterloo, Waterloo, Ontario N2L 3G1, Canada}

\begin{abstract}
{We revisit the ``field entanglement in non-inertial frames" scenario extensively studied in previous literature and demonstrate that the geometric discord, a proposed measure of generalized quantum correlations, decays to zero in the infinite acceleration limit. This is in contrast with previous research showing that the acceleration-induced degradation of quantum discord was not strong enough to extinguish discord in this limit.  We argue that our finding has two opposing but non-contradicting implications. On the one hand the usable quantum correlations in the large acceleration regime appear severely limited for any protocols making use of geometric
discord as a figure of merit. On the other hand our result indicates, in corroboration with other recent work, that the geometric discord is not a faithful measure of quantum correlations, especially in the case of continuous variable systems due to the existence of states which have finite quantum discord but are nevertheless infinitesimally close (in the Hilbert-Schmidt norm) to a zero-discord state.}

\end{abstract}


\pacs{03.67.-a, 04.62.+v}

\maketitle

\newcommand{\qr}{q_\text{R}}

\section{Introduction}

In recent years there has been a growing level of interest in the field of relativistic quantum information. Some of the seminal work performed in this regard has been the study of entanglement between quantum field modes as observed by relatively accelerating observers \cite{Robb,Robb2}, extraction of field entanglement using the expansion of the universe \cite{expansion,expansion2}, and the study of entanglement in black hole spacetimes \cite{blackhole,blackhole2}. Research of this sort is useful and interesting both from the viewpoint of obtaining  a fundamental understanding of nature as well as from the practical side of things. Seeing as we live in a relativistic world it makes sense to understand quantum information in this regime, and not just in the Newtonian limit. Additionally it may help to guide us in the implementation of quantum circuits and the quantification of both the resources required to overcome, and the resources provided by, relativistic effects.

The flat-space scenario that we will consider is the same as that considered in \cite{Robb,Robb2}. Alice is an inertial observer who shares a maximally entangled state of a quantum field with a non-inertial observer named Rob. It is found that the acceleration of Rob induces a degradation of the entanglement such that, for a bosonic field, the entanglement decays to zero in the limit of infinite acceleration. Of particular interest in this article however is the quantification of quantum discord rather than of entanglement. Quantum discord, originally introduced in \cite{quantdis}, is a measure of quantum correlations that extends beyond entanglement. This quantity is useful in the case of mixed states (which our state of interest \(\rho_{AR}\) is an example of), and generally it is possible to have separable mixed states that nevertheless have nonzero discord. Indeed it was shown in \cite{Datta} that the decay of quantum discord \(D_1(\rho_{AR})\) of a bosonic field in our non-inertial scenario decays to a finite value in the infinite acceleration limit, despite the fact that the entanglement decays to zero. Since discord has been shown to be a usable resource in mixed state quantum computation and communication \cite{Datta2,Modi11,Gu,Madhok1}, this result indicates the possibility of Rob being capable of performing quantum computational tasks even in the case of large accelerations. A similar study was performed for a fermionic field \cite{Wang} in which case similar behaviour is witnessed. However entanglement in the fermionic case in this limit does not vanish either, and so the utility of discord as a resource remains unclear.

{
Our goal in this article is to contrast the quantum discord \(D_1\) as computed in \cite{Datta} with the geometric discord \(D_G\) in the same Alice-Rob scenario, which in fact vanishes in the infinite acceleration limit. The geometric discord \cite{geodis,geodis2} is an alternative definition of total quantum correlations that is defined as (the square of) the Hilbert-Schmidt distance between the state in question and the nearest state of zero quantum discord. Both the quantum and geometric discords have been given operational significance (see for example \cite{statemerging} and \cite{stateprep} respectively).
}


Our result is peculiar in the sense that we have a state which has finite quantum discord but is nevertheless infinitesimally close (in the Hilbert-Schmidt norm) to a zero-discord state. This behaviour was also observed in \cite{CVdiscord}, and is seen there to be a result of considering states on infinite dimensional Hilbert spaces (which will appear when studying field states as seen from accelerated observers) that have unbounded energies (which of course these states do in the infinite acceleration limit). In fact similar behaviour is observed for entanglement in infinite dimensional systems as well, where it is known that unless one bounds the mean energy the states of infinite entanglement entropy form a trace-norm dense set \cite{CVentangle}. Our discovery of vanishing geometric discord for large accelerations is commensurate with these findings. {Quite generally therefore one must keep in mind questions of dimensionality and energy when attempting to quantify correlations, especially in the case of infinite dimensions \cite{CVquantdiscord,CVdiscord,Disc1,Disc2}. Here we will discuss two implications of the discrepancy between the quantum and geometric discords for large accelerations. The first is that usable quantum operations are perhaps somewhat more limited in this regime than previously thought, and the second being that the geometric discord is not a faithful measure of quantum correlations in the case of infinite dimensions.}



The majority of work on quantum discord has focused on finite dimensional systems and their extension to infinite dimensions is unclear.
As this paper was being completed however, Tufarelli et al. \cite{Adesso} demonstrated -- by considering a closed form solution of the geometric discord for \(2 \times d\) dimensional systems including the case  \(d=\infty\) --  that geometric discord is the essential operative ingredient for remote quantum state preparation \cite{stateprep}, a variant of the quantum teleportation protocol, even in the infinite-dimensional case. Such a result has obvious implications for the Alice-Robb scenario considered here. {In particular it indicates, given the vanishing of both entanglement and geometric discord for large accelerations, that there is at least a significant set of quantum operations that are restricted in this limit. In particular, this statement is at the very least true for quantum communication protocols: while the teleportation fidelity vanishes due to the lack of entanglement we see via vanishing geometric discord that remote state preparation becomes impossible as well. Interestingly it has very recently been pointed out that geometric discord is also the figure of merit for non-locality as quantified by the global change of state induced by local unitary evolution \cite{nonlocal}, however it is unclear if the conclusions of this work extend to infinite-dimensional systems as the one considered here.

{
One may be tempted to make a claim stronger than that given above and declare that no usable quantum correlations remain in a state of zero geometric discord. An innocent argument to be made for this is that two states with vanishing Hilbert-Schmidt norm should be indistinguishable from any measurement procedure that can be performed on them. While this argument may hold for finite-dimensional systems, it is easily seen not to be the case for infinite dimensions. We demonstrate this explicitly in section \ref{appendix}, where two states with vanishing Hilbert-Schmidt distance (the same states considered in section \ref{bosonic}) are shown to have divergingly different expectation values of joint observables. Furthermore we demonstrate in section \ref{appendix} that even though the Hilbert-Schmidt (2-norm) distance vanishes, and therefore all higher-norm distances, the trace (1-norm) distance remains finite. This further indicates that in infinite dimensions the geometric discord is not a faithful quantifier of quantum correlations. We therefore cannot conclude that for large acceleration all quantum correlations reach negligible levels, rather this is only so for a subset of such correlations as discussed above. However even given that quantum correlations survive in this limit it remains unclear if one could practically access them for use. We discuss this point further in section \ref{appendix}.}

{We note that very recently the geometric discord has been revealed to have unappealing properties independent of the question of dimensionality \cite{piani}. Our findings are independent and commensurate with the results presented in this work regarding the notion that geometric discord is an inadequate measure of general quantum correlations.
}

In this paper,  we also compute, for completeness, the geometric discord  for fermionic fields. In this case we find that it does not decay to zero in the infinite acceleration limit. Of course there is no reason to suspect that it should because Rob's Hilbert space is no longer of infinite dimension.

Our presentation will proceed as follows. In Sect. \ref{acceleration} we will discuss the problems with, and the justification for using, the Unruh modes that we employ here. We will also complete our discussion of our state \(\rho_{AR}\), for the cases of both a bosonic and a fermionic field. In Sect. \ref{discords} we will give a brief introduction to the quantum and geometric discords and define their form. In Sect. \ref{bosonic} we will present the primary result of this article, namely the result of vanishing geometric discord in the case of a bosonic field. In Sect. \ref{fermionic} we will briefly give the result of similar calculations performed this time using a fermionic field, and then {in section \ref{appendix} we demonstrate that the result of vanishing geometric discord in infinite dimensions does not imply absence of quantum correlations.   Sect. \ref{conclusions} contains some concluding remarks.}

\section{An Accelerated Observer} \label{acceleration}

We will be interested in the flat-space scenario considered by \cite{Robb,Robb2} in which one observer (Alice) is inertial and the other (Rob) is undergoing uniform acceleration. It was shown in these articles that if Alice and Rob share a maximally entangled state in terms of Minkowski modes then the degree of entanglement is degraded due to the acceleration of Rob. Physically this can be interpreted as being due to the decohering effect of Unruh radiation as perceived by Rob. In the case of a bosonic field it was found that the entanglement vanishes to zero in the infinite acceleration limit. For a fermionic field we similarly see a degradation of entanglement, but in this case it asymptotes to a finite value in the infinite acceleration limit. To describe a field from Rob's perspective we must transform to an appropriate set of coordinates, namely the Rindler coordinates \((\tau, \chi)\). If Rob is accelerating in the Minkowski direction \(z\) with proper acceleration \(a\) then these coordinates are defined implicitly by
\begin{align}
    t=a^{-1}e^{a\chi} \sinh a\tau, \;\;\; z=a^{-1}e^{a\chi}\cosh a\tau
\end{align}
Note that these coordinates only cover the right Rindler wedge to which Rob is confined. A straightforward analytic continuation lets us include coordinate patches for the rest of spacetime. This includes the left Rindler wedge in which we can define a duel accelerating observer, imaginatively named antiRob. The coordinate patch for the left wedge takes the form
\begin{align}
    t=-a^{-1}e^{a\chi} \sinh a\tau, \;\;\; z=-a^{-1}e^{a\chi}\cosh a\tau
\end{align}
In their respective wedges Rob and antiRob travel along the paths \(\chi=0\), and \(\tau\) is their proper time.

These new coordinates allows us to perform a Bogoliubov transformation between the Minkowski modes of a field (eigenvectors with respect to Lie transport along \(t\)) and Rindler modes (eigenvectors with respect to Lie transport along \(\tau\)) \cite{bd}. The Rindler modes as described by Rob and antiRob together form a complete basis with which we can expand the Minkowksi modes. Therefore an arbitrary state as perceived by Alice can be represented in this basis as well. However we know that Rob and antiRob are causally disconnected, which means that a state as perceived by Rob cannot have any contributions from the antiRob modes. To obtain the part of the state common to Alice and Rob we must therefore trace out the left Rindler wedge in which antiRob lives. In general the resulting reduced state will be mixed, meaning that we must describe it with a density matrix \(\rho_{AR}\). It is this state that we will be concerned with.

As has been done extensively in the past, and what we will do as well, is to simplify our work by replacing an excited Minkowski mode as perceived by Rob with what's known as an Unruh mode \cite{Bruschi2010}. Such modes form an alternative complete set of solutions from the Minkowskian perspective. These are a particularly convenient set of solutions to utilize because they map to single frequency Rindler modes while nevertheless sharing the vacuum of Minkowski modes. While the use of Unruh modes is not without its problems they can still be useful in making qualitative predictions. This is especially true in the infinite acceleration limit, which is where our focus is here. We now proceed to give a more complete description of this procedure and the justification of its use.

We will consider the following family of bipartite, maximally entangled states of a scalar field prepared from the inertial perspective commonly used in the literature on relativistic quantum information \cite{Bruschi2010},
\begin{equation}\label{state}
\ket{\Psi}_{\text{AR}}=\frac{1}{\sqrt2}\left(\ket{00}+\ket{11}\right),
\end{equation}
where the states $\ket1$ belonging to Rob's subsystem are Unruh modes of a given Rindler frequency $\omega$ as seen by accelerated observers with proper acceleration $a$.

An arbitrary Unruh mode for a given acceleration has the form \cite{Bruschi2010},
 \begin{align}C_\omega=q_\text{L}C_{\omega,\text{L}}+\qr C_{\omega,\text{R}}\label{umodes},\end{align}
where $\vert q_\text{L}\vert^2+\vert\qr\vert^2=1$, and
\begin{eqnarray}\label{bogoboson}
 C_{\omega,\text{R}}&=&\cosh r_\omega\, a_{\omega,\text{I}} - \sinh r_\omega\, a^\dagger_{\omega,\text{II}},\\*
 C_{\omega,\text{L}}&=&\cosh r_\omega\, a_{\omega,\text{II}} - \sinh r_\omega\, a^\dagger_{\omega,\text{I}}, \end{eqnarray}
where \(r_\omega\) is defined by $\tanh r_\omega=e^{-\pi c\,\omega/a}$ and $a_{\omega,\text{I/II}}, \, a^\dagger_{\omega,\text{I/II}}$ are Rindler particle operators for the scalar field in the left and right spacetime wedges respectively.

Note that the use of Unruh modes is not free from problems: due to the their ill-localisation these modes cannot be completely measured, and at best can only be approximately determined by means of a localised measurement. Additionally the Unruh modes as seen by any inertial observer are highly oscillatory   near the acceleration horizon. This makes them bad candidates for physically feasible states. In this paper we are choosing a different Unruh mode for each value of the proper acceleration, so as to keep the Unruh to Rindler change of basis always simple. This simplification is common in the literature (although in the papers preceding ref. \cite{Bruschi2010} it was sometimes used in a very obscure way) and has been employed extensively in recent years \cite{Montero2012,Montero2011a,Wang2011,Montero2011b,Friis2011,Montero2010}. We are interested in using these modes in order to contrast the fundamental results obtained here with those in previous literature that made use of these modes \cite{Datta}.

Notice that the most general state of positive Minkowski frequency can always be decomposed into a linear combination of Unruh creation operators in the left and right wedges, so by studying these modes we are analysing the behavior of a convenient complete basis of solutions of the field equations that can span any physical state.

As with other work on relativistic quantum information employing Unruh modes (e.g. \cite{Robb, Robb2, Datta, Smith:2011mg} among others)
we obtain a 1-parameter family of solutions depending on $r$, which is related not only to the acceleration parameter but also to the choice of state.
The use of Unruh-modes to construct the entangled state (\ref{state}) yields two nice features: this one-parameter family of states are maximally entangled in the inertial limit and translate to single frequency entangled states from the point of view of constantly accelerated observers.
Of course, translating the results obtained in this scenario to carry out concrete experiments is not straightforward and other approaches should be used (such as localized projective measurements \cite{localPVM} or particle detector models \cite{Ostapchuk:2011ud}).

 Nevertheless,   the use of Unruh modes to extract conclusions regarding fundamental physics is justified due to the following desirable features.
\begin{itemize}
\item They are purely positive frequency linear combinations of Minkowski modes.
\item The vacuum for the Unruh modes is the same as for Minkowski monochromatic modes.
\item  They have a sharp frequency when translated to the Rindler basis.
\item  They yield a complete set of orthonormal solutions of the Minkowski coordinate field equations.
\end{itemize}

Furthermore, while the results we obtain map out the behaviour of different states as the parameter $r$ changes, the features noted above ensure that  in the inertial limit (or small-$r$ limit -- which is also the small acceleration limit)  they will describe the behaviour of maximally entangled physical states and in the large-$r$ limit (which is also the large acceleration limit) they will also yield the correct behaviour for physical states. So although   quantitatively some details for intermediate values of $r$ could differ for physical
states, our results qualitatively describe the behaviour for the relevant states as a function of the acceleration and  our main result
-- the vanishing of geometric discord
 in the infinite acceleration limit -- remains valid for both Unruh modes and physical states.

Therefore, for all these reasons, the field excitations in this work will be considered for convenience as Unruh modes $\ket{1}=C^\dagger_\omega\ket{0}$. All these states have an implicit dependence on Rob's acceleration $a$ when expressed in the Rindler basis through a parameter $r_\omega$ defined by $\tanh r_\omega=e^{-\pi c\,\omega/a}$.

We will restrict our considerations to a particular choice of Unruh modes $\qr=1\Rightarrow q_L=0$. This is the regime known as the single mode approximation \cite{Bruschi2010}. Restriction to this regime allows us to most straightforwardly demonstrate our results, which
 trivially carry over to any other choices of Unruh modes. Previous work  analyzing discord in field states from non-inertial perspectives \cite{Datta} were also limited to this regime, thereby making comparisons simpler.

A  Minkowski vacuum
state is defined as the absence of any particle excitation in any of
the modes
\begin{equation}
|0\rangle ^{\mathcal{M}}=\prod_{k}|0_{k}\rangle ^{\mathcal{M}},
\end{equation}
and can be expressed in terms of a product of two-mode squeezed states
of the Rindler vacuum
\begin{align}\label{eq:vacuum}
    \ket{0_k}^\mathcal{M}&=\frac{1}{\cosh r}\sum_{n=0}^\infty \tanh^n r \, \ket{n_k}_I \ket{n_k}_{II} \\
    \cosh r &=(1-e^{-2\pi \Omega})^{-1/2}, \;\;\; \Omega=|k|c/a \nonumber
\end{align}
where  $\left| n_{k}\right\rangle _{I}$ and $\left|
n_{k}\right\rangle _{II}$ refer to the mode decomposition in
region I and II, respectively, of Rindler
space. Each vacuum Minkowski mode $j$ has a Rindler mode expansion given by Eq. (\ref{eq:vacuum}).
Additionally, the one-particle Unruh mode \(\ket{1_K}^{\mathcal{U}} \equiv C_k^\dagger \ket{0}^{\mathcal{M}}\) using \(q_R=1\) has a Rindler expansion given by
\begin{equation}
\left| 1_{k}\right\rangle ^{\mathcal{U}}=\frac{1}{\cosh ^{2}r}%
\sum_{n=0}^{\infty }\tanh ^{n}r\,\sqrt{n+1}\left|
(n+1)_{k}\right\rangle _{I}\left| n_{k}\right\rangle _{II},
\notag
\end{equation}%
We can rewrite Eq.(\ref{state}) in terms of Minkowski modes for
Alice and Unruh modes for Rob. Since Rob is causally
disconnected from region II, we must trace over the states in this
region, which results in a mixed state
\begin{align}\label{denstate}
    \rho_{AR}=\frac{1}{2\cosh^2r}\sum_{n=0}^\infty \tanh^{2n}r \, \rho_n
\end{align}
where
\begin{align}
    \rho_n=&\ket{0,n}\bra{0,n}+\frac{\sqrt{n+1}}{\cosh r}\ket{0,n}\bra{1,n+1} \\ \nonumber
    &+\frac{\sqrt{n+1}}{\cosh r}\ket{1,n+1}\bra{0,n}+\frac{n+1}{\cosh^2 r}\ket{1,n+1}\bra{1,n+1}
\end{align}

We will find it convenient to rewrite this state in the following form,
\begin{align} \label{bosonstate}
    \rho_{AR}=\frac{1-t^2}{2}(\ket{0}\bra{0}\otimes M_{00}+\ket{1}\bra{1}\otimes M_{11} \nonumber \\
    \ket{0}\bra{1}\otimes M_{01}+\ket{1}\bra{0}\otimes M_{10})
\end{align}
where \(t \equiv \tanh r\) and the matrices on Rob's Hilbert space are
\begin{align} \label{Ms}
    &M_{00}=\sum_{n=0}^\infty t^{2n}\ket{n}\bra{n},  \\
    M_{11}=(1&-t^2)\sum_{n=0}^\infty (n+1)t^{2n}\ket{n+1}\bra{n+1}, \nonumber \\
    M_{01}=\sqrt{1-t^2}&\sum_{n=0}^\infty \sqrt{n+1} \, t^{2n}\ket{n}\bra{n+1}, \; M_{10}=M_{01}^\dagger \nonumber
\end{align}

For the Dirac case a similar analysis \cite{Robb2} implies that  the Dirac Minkowski fields expanded in terms of Unruh modes yields
\begin{align}
    \ket{0_k}^{\mathcal{U}} &= \cos r \, \ket{0_k}_I \ket{0_{-k}}_{II}+\sin r \, \ket{1_k}_I \ket{1_{-k}}_{II}\\
    \ket{1_k}^{\mathcal{U}} &= \ket{0_k}_I \ket{1_{-k}}_{II}
\end{align}
where $\ket{0}^U = \ket{0}^M$  since the Unruh modes are just linear combinations  of purely positive frequency Minkowski modes.
Following the same procedure as that performed for the bosonic field, the Alice-Rob reduced density matrix takes the form \cite{Robb2}
\begin{align}\label{fermistate}
    \rho_{AR}=\frac{1}{2}[\cos^2 r\ket{00}\bra{00}+\sin^2 r\ket{01}\bra{01} \nonumber \\
    +\cos r(\ket{00}\bra{11}+\ket{11}\bra{00})+\ket{11}\bra{11}]
\end{align}

\section{The Quantum and Geometric Discords} \label{discords}

Quantum discord is a measure of quantum correlations. It coincides with entanglement for pure states, but in the case of mixed states represents a broader measure of quantum correlations than does entanglement, as it can persist even for separable states \cite{Modi11}. Quantum discord is thought to be crucial in quantum algorithms that outperform their classical counterparts but where little to no entanglement is present \cite{Datta2,Modi11,Gu,Madhok1}. It is therefore a matter of important interest both from a foundational and practical point of view to understand quantum discord in physical systems. In particular it is of interest in relativistic settings as space-time horizons such as one that an observer in Rindler space experiences often have the effect of turning initially pure states into mixed states as in the case considered here. In this regime therefore discord might capture non-classical behaviour that is not encompassed by entanglement.

In previous work detailing quantum correlations in relativistic settings \cite{Datta,Wang} the original definition of discord  \cite{quantdis} was analyzed. In this definition, the discord of a state \(\rho_{AB}\) composed of two subsystems \(A\) and \(B\)  is given by
\begin{align} \label{discord1}
    D^A_1(\rho_{AB}) \equiv S(\rho_A)+\min_{\{\Pi_A\}}S(\rho_B|\{\Pi_A\})- S(\rho_{AB})
\end{align}
where \({S(\rho)=-\text{Tr}(\rho \log_2(\rho))}\) is the Von Neumann entropy and
\begin{align} \label{VNentropy}
    S(\rho_B|\{\Pi_A\})=-\sum_{a} p_a S(\rho_b|\Pi_a)
\end{align}
is the conditional Von Neumann entropy; the Von Neumann entropy of subsystem B after the projective measurement \(\{\Pi_a\}\) has been performed on system A. Here \(p_a=\text{Tr}(\Pi_a \otimes I \rho_{ab} \Pi_a \otimes I)\) is the probability of the measurement resulting in outcome \(\Pi_a\).
The superscript \(A\) in Eq. (\ref{discord1}) serves to identify that the measurement is being performed on subsystem \(A\). In general the discord is asymmetric upon this interchange
\begin{align}
    D^A(\rho_{AB}) \ne D^B(\rho_{AB})
\end{align}
where the subscript 1 is omitted because this relationship also holds for the geometric discord, which will now be introduced.


The geometric discord is a metric based measure of quantum correlations. It is measured by the distance in state space from the state in question to the nearest state with zero discord. Such zero discord states are known as classical-quantum (quantum-classical) states when considering the case that measurement is performed on the subsystem \(A\) (\(B\)). The state-space distance is measured by the square of the Hilbert-Schmidt distance. The geometric discord is given by

\begin{align}\label{defdisc}
    D^A_G(\rho_{AB}) \equiv \min_{\chi \in \mathcal{C}}||\rho_{AB}-\chi||
\end{align}
where \(\mathcal{C}\) is the set of classical-quantum states and
\begin{align}
    ||\rho-\chi|| \equiv \text{Tr}((\rho-\chi)^2)
\end{align}
is the squared Hilbert-Schmidt norm. The reader should note that the geometric discord has been shown to be an experimentally accessible quantity \cite{Adesso,Jin}.
It was also shown in \cite{geodis2} that the geometric discord is equivalent to a somewhat more workable form given by
\begin{align} \label{geodiseqn}
    D^A_G(\rho)=\min_{\{\Pi_a\}}||\rho-\rho'||=\min_{\{\Pi_a\}}\text{Tr}((\rho-\rho')^2)
\end{align}
where \(\rho'\) is the state after the projective measurement \(\{\Pi_a\}\) has been performed on subsystem \(A\) as given by
\begin{align}
    \rho' &\equiv \sum_a(\Pi_a \otimes I)\rho \,(\Pi_a \otimes I)
\end{align}
This second form of the geometric discord is the one that we will be using to perform our calculations.

\section{Discord and Bosonic Fields} \label{bosonic}

We continue now to the computation of geometric discord for the Alice-Rob system described above, where here we focus on the case of a bosonic field. It is in this case we find that in the infinite acceleration limit \(D_G(\rho_{AR})\) decays to zero, despite the fact that the quantum discord \(D_1(\rho_{AR})\) limits to a finite value as shown in \cite{Datta}. As explained in the introduction, this peculiar behavior is actually what should be expected in the case of infinite dimensional systems with unbounded energy.

Computing the geometric discord \(D_G\) for this system is straightforward but somewhat tedious. We will outline the steps to be taken in this calculation and present the end result, but first it may be useful to give a much faster calculation that illustrates our primary finding: that \(D_G\) vanishes in the large acceleration limit. In this simplified calculation we won't worry about the minimization over projective measurements but rather just pick one that makes the calculation easy. This will give us an upper bound on \(D_G\), which we will see also vanishes for infinite accelerations, thus proving our result. Indeed, the fact that this occurs for \emph{any} projective measurement on Alice's system is an interesting result in its own right.

The measurement to be taken is effectively over a single qubit system and can therefore be parameterized by the unit vector \(\mathbf{x}=(x_1,x_2,x_3)\), \(x_1^2+x_2^2+x_3^2=1\), where the projectors are given by
\begin{align} \label{projectors}
    \Pi_\pm = \frac{1}{2}(I\pm \mathbf{x \cdot {\sigma}})
\end{align}
where \(\mathbf{\sigma}=(\sigma_1,\sigma_2,\sigma_3)\) are the Pauli matrices and \(I\) is the \(2 \times 2\) identity. As we will see, the geometric discord turns out to be independent of \(x_1\) and \(x_2\) and is minimized when \(x_3=0\), which is the same projective measurement found to minimize the quantum discord in \cite{Datta}.

First, let us compute an upper bound by choosing our measurement such that \(x_3=1\) (and thus \(x_1=x_2=0\)); this corresponds to the projectors \(\Pi_+=\ket{0}\bra{0}\) and \(\Pi_-=\ket{1}\bra{1}\). When this choice is used we will label the resulting geometric discord \(D_G|_{x_3=1}\), and by definition it must be greater than or equal to the true geometric discord \(D_G\). Recalling Eq. (\ref{geodiseqn}), we have
\begin{align}
    D_G \leq D_G \big |_{x_3=1}=\text{Tr}((\rho_{AR}-\chi)^2)
\end{align}
where \(\chi\) is the state after the measurement \(\{\ket{0}\bra{0}, \ket{1}\bra{1}\}\) has been performed. Clearly this acts to eliminate the off-diagonal terms from \(\rho_{AR}\),
\begin{align} \label{postmeasuredboson}
    \rho_{AR} \rightarrow \chi &=\sum_{a=0,1}(\ket{a}\bra{a} \otimes I_R)\rho_{AR} \,(\ket{a}\bra{a} \otimes I_R) \nonumber \\
    &=\frac{1-t^2}{2}(\ket{0}\bra{0} \otimes M_{00}+\ket{1}\bra{1} \otimes M_{11})
\end{align}
The operator \(\rho_{AR}-\chi\) will therefore be the purely off-diagonal part of \(\rho_{AR}\), and upon squaring will again be diagonal and of the form
\begin{align}
    (\rho_{AR}-\chi)^2=\frac{(1-t^2)^2}{4}(&\ket{0}\bra{0}\otimes M_{01}M_{10} \nonumber \\
    +&\ket{1}\bra{1}\otimes M_{10}M_{01})
\end{align}

Note that \(M_{01}M_{10}\) is diagonal, and to finish the calculation we need to compute the trace \(\text{Tr}(M_{01}M_{10})\). We will also find it necessary later to compute the traces of other products of \(M\) matrices, and so we will list them now. The nonzero traces of such products are easily found via geometric series,
\begin{align} \label{Mtraces}
    &\text{Tr}(M_{00}^2)=\frac{1}{1-t^4}, \; \; \text{Tr}(M_{11}^2)=\frac{1+t^4}{(1+t^2)^2(1-t^4)}, \nonumber \\
    &\text{Tr}(M_{00}M_{11})=\text{Tr}(M_{11}M_{00})=\frac{t^2}{(1+t^2)(1-t^4)},  \nonumber \\
    &\text{Tr}(M_{01}M_{10})=\text{Tr}(M_{10}M_{01})=\frac{1}{(1+t^2)(1-t^4)}
\end{align}
With this, we conclude
\begin{align} \label{upperbound}
    D_G \big |_{x_3=1}=\frac{(1-t^2)^2}{2}\text{Tr}(M_{01}M_{10})=\frac{1-t^2}{2(1+t^2)^2}
\end{align}
which indeed decays to zero in the infinite acceleration limit \(t \rightarrow 1\). Since \(D_G \leq D_G|_{x_3=1}\), this simple calculation demonstrates that \(D_G \rightarrow 0\) in this limit as well.

Although the previous calculation demonstrates our primary result, for completeness let us also include the full result of how \(D_G\) behaves as a function \(t\). While doing this is much more tedious than the previous calculation, it is nevertheless straightforward and we will outline here the steps to be taken. Using the projectors in Eq. (\ref{projectors}) the geometric discord is given by the minimization of \(\text{Tr}((\rho_{AR}-\rho_{AR}')^2)\) over the parameters \((x_1,x_2,x_3)\). The post-measured state \(\rho_{AR}'\) is
\begin{align}
    \rho_{AR}'=\sum_{a=\pm}(\Pi_a \otimes I_R)\rho_{AR}(\Pi_a \otimes I_R)
\end{align}
and we will find it useful to consider this in the form
\begin{align}
    \rho_{AR}'=\sum_{a=\pm} p_a \Pi_a \otimes \rho_{R|a}
\end{align}
where \(\rho_{R|a}\) is the post-measured state of Rob's reduced system conditioned on the outcome \(a\),
\begin{align}
    \rho_{R|a} \equiv \text{Tr}_{A}((\Pi_a \otimes I_R)\rho_{AR}(\Pi_a \otimes I_R))/p_a
\end{align}
and \(p_a\) is the probability of measurement outcome \(a\),
\begin{align}
    p_a \equiv \text{Tr}((\Pi_a \otimes I_R)\rho_{AR}(\Pi_a \otimes I_R))
\end{align}

The projectors in Eq. (\ref{projectors}) can be written as
\begin{align} \label{projectors2}
    \Pi_\pm=\frac{1}{2}[&(1\pm x_3)\ket{0}\bra{0}+(1\mp x_3)\ket{1}\bra{1} \nonumber \\
    &\pm(x_1-i x_2)\ket{0}\bra{1}\pm (x_1+i x_2)\ket{1}\bra{0}] \nonumber
\end{align}
from which it is easily seen that the outcome probabilities equate to \(p_\pm=1/2\) and that
\begin{align}
    \rho_{R|\pm}=\frac{1-t^2}{2}\tilde{\rho}_\pm
\end{align}
where
\begin{align}
    \tilde{\rho}_\pm \equiv &(1\pm x_3)M_{00}+(1\mp x_3)M_{11} \nonumber \\
    &\pm(x_1-i x_2)M_{01}\pm (x_1+i x_2)M_{10}
\end{align}
Note that the result \(p_\pm=1/2\) is entirely expected since it is easily seen that Alice's reduced state is maximally mixed, \(\rho_A=\text{Tr}_R \rho_{AR}=I/2\), and so the outcomes of any projective measurement must have probabilities \(1/2\).

From these preliminaries it is simple to show that the quantity we need to evaluate is given by
\begin{align}
    \text{Tr}((\rho_{AR}-\rho_{AR}')^2)=&\frac{(1-t^2)^2}{4}[\text{Tr}(X_{00}^2) \nonumber \\
    &+\text{Tr}(X_{11}^2)+2\text{Tr}(X_{01}X_{10})]
\end{align}
where
\begin{align}
    &X_{00} \equiv M_{00}-((1+x_3)\tilde{\rho}_+ +(1-x_3)\tilde{\rho}_-)/4 \nonumber \\
    &X_{11} \equiv M_{11}-((1-x_3)\tilde{\rho}_+ +(1+x_3)\tilde{\rho}_-)/4 \nonumber \\
    &X_{01} \equiv M_{01}-(x_1-ix_2)(\tilde{\rho}_+-\tilde{\rho}_-)/4 \nonumber \\
    &X_{10} \equiv M_{10}-(x_1+ix_2)(\tilde{\rho}_+-\tilde{\rho}_-)/4
\end{align}
The traces \(\text{Tr}(X_{00}^2)\), \(\text{Tr}(X_{11}^2)\) and \(\text{Tr}(X_{01}X_{10})\) are given by linear combinations of the traces presented in Eq. (\ref{Mtraces}) such that the above equation reduces to
\begin{align}
     \text{Tr}&((\rho_{AR}-\rho_{AR}')^2)=\frac{(1-t^2)^2}{8}[(1-x_3^2)(\text{Tr}(M_{00}^2) \nonumber \\
     &+\text{Tr}(M_{11}^2)-2\text{Tr}(M_{00}M_{11}))+2(1+x_3^2)\text{Tr}(M_{01}M_{10})] \nonumber \\
     &=\frac{1-t^2}{4(1+t^2)^3}(2+t^2+x_3^2t^2)
\end{align}
where Eq. (\ref{Mtraces}) was used in the last equality. Note that this quantity is independent of \(x_1\) and \(x_2\) and that if we set \(x_3=1\) we obtain the same result as Eq. (\ref{upperbound}), as we must. In any case, this is clearly minimized when \(x_3=0\), and we have
\begin{align}
    D_G(\rho_{AR})=\frac{(1-t^2)(2+t^2)}{4(1+t^2)^3}
\end{align}
Note that the quantum discord $D_1(rho_{AR})$ also takes its minimum at $x_3 =0$, as found by \cite{Datta}. Unlike the quantum discord however, the geometric discord approaches zero as \(t \rightarrow 1\).

It is also interesting to point out that the result \(D_G(\rho_{AR})\rightarrow 0\) as \(t\rightarrow 0\) applies also to the case when the measurement is performed on Rob's system rather than on Alice's. In general the definition of geometric discord as given by Eq. (\ref{geodiseqn}) is not symmetric between the subsystems, and so we would generally expect \(D_G(\rho_{AR})\) to depend on who's system we are performing a measurement on. This is also true of the quantum discord and the results found by Datta \cite{Datta} were for the case in which Alice's system is the one that is measured, as is also the case that we have examined here. Unfortunately it is unknown how to perform the same calculation when Rob's system is measured. The difficulty arises when attempting to perform the minimization over projective measurements, which in Rob's case is a minimization over an infinite set of parameters rather than the two parameters encountered on Alice's side. Thus we seem stuck with a one-sided view of the discord in our considered scenario.

However, we are always free to compute an upper bound by making a choice of measurement, as was done above to compute \(D_G|_{x_3=1}\). This can just as easily be done when the measurement is performed over Rob's system, in which case we choose the set of projectors \(\{\ket{n}\bra{n}\}\) over all \(n\). In this case the geometric discord satisfies \(D_G^{(R)}(\rho_{AR}) \leq \text{Tr}((\rho_{AR}-\chi_R)^2)\), where we use the notation \(D_G^{(R)}(\rho_{AR})\) to indicate that now the measurement is over Rob's system. We now observe that the post-measured state \(\chi_R\) is equivalent to the post-measured state found above, Eq. (\ref{postmeasuredboson})! That is,
\begin{align}
    \rho_{AR} \rightarrow \chi_R&=\sum_{n=0}^\infty (I_A \otimes \ket{n}\bra{n})\rho_{AR}(I_A \otimes \ket{n}\bra{n}) \nonumber \\
    &=\frac{1-t^2}{2}(\ket{0}\bra{0} \otimes M_{00}+\ket{1}\bra{1} \otimes M_{11})
\end{align}
The subsequent calculation will therefore follow exactly as above, and we conclude that \(D_G^{(R)}(\rho_{AR}) \rightarrow 0\) as \(t\rightarrow 1\). That is, the geometric discord vanishes in the infinite acceleration limit both when the measurement is performed over Alice's system \emph{and} when performed over Rob's system. The reader should note that such an upper bound calculation can also be performed for the quantum discord \(D_1^{(R)}(\rho_{AR})\), but the result is uninformative as the upper bound simply approaches the finite value \(\approx 0.85\).

\section{Discord and Fermionic Fields} \label{fermionic}

It has been shown previously that when Rob and Alice share a maximally mixed Minkowski state, Eq.(\ref{state}), of a Dirac field that entanglement \cite{Robb2} and discord when Alice does the measurement \cite{Wang} decrease with increasing acceleration, but that both quantities remain finite in the infinite acceleration limit. We calculate here geometric discord considering cases where both Alice and Rob perform the measurement. Geometric discord decreases as acceleration increases but remains finite in the infinite acceleration limit similar to \(D_1\) as can be seen in figure \ref{fermigraph}. It is not surprising to find that the geometric discord limits to a finite value in this case, because we no longer have the infinite dimensional Hilbert space required to see such a characteristic.

While the calculations in this section are straightforward they are somewhat tedious. For this reason we outline the calculation of the discord in the case Alice where does the measurement, but simply present the results for the case when Rob
does the measurement. Because the Hilbert space is now finite it is straightforward  to complete these calculations and optimizations by direct calculation.
\begin{figure}
\includegraphics[width=0.5\textwidth]{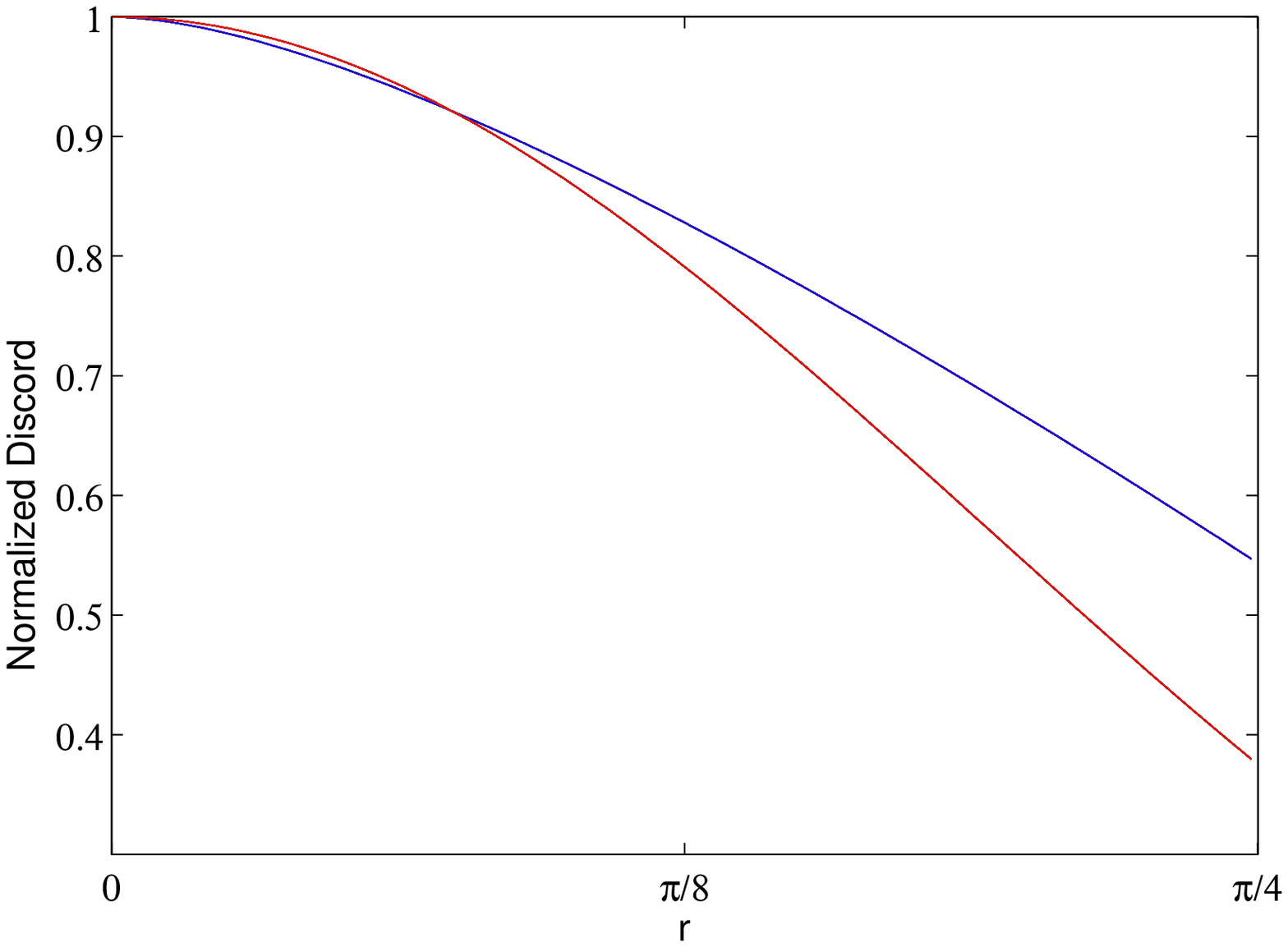}
\includegraphics[width=0.5\textwidth]{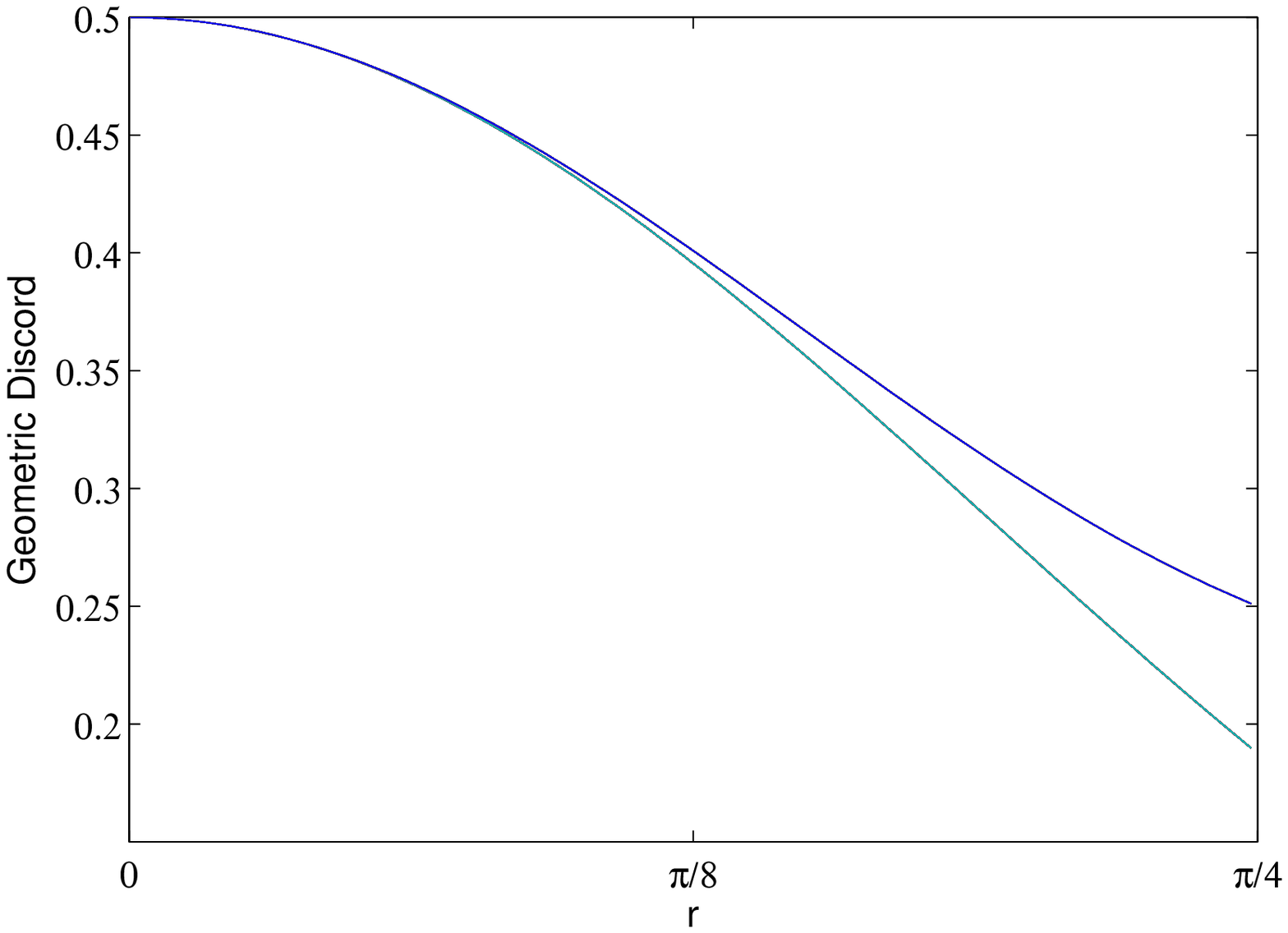}
\caption{(Color online) Discord of the Alice-Rob shared state Eq. (\ref{fermistate}) of a Dirac field.
The top figure compares the quantum discord \(D_1\) (blue, upper curve) with twice the geometric discord \(2D_G\) (red, lower curve) as a function of $r=\operatorname{atan}(e^{-\pi c\,\omega/a})$ in the case that Alice performs the measurement. The bottom figure compares \(D_G\) between the cases in which Rob performs the measurement (blue, upper curve) and when Alice performs the measurement (green, lower curve).}
\label{fermigraph}

\end{figure}

In calculating the geometric discord we follow here an approach similar to that used in the bosonic case.
In the case that Alice does the measurement her density matrix is the maximally mixed state. Therefore the probability of measuring any outcome is 1/2 just as in the bosonic case. In fact it is easy to construct analogues of the \(M_\sigma\) matrices
\begin{align}
    R_{00} &= \sum\limits_{n=0}^{1} \cos^2 r \tan^{2n} r \ket{n}\bra{n} \\
    R_{11} &= \ket{1}\bra{1} \\
    R_{10} &= R^\dagger_{01}= \cos r \ket{1}\bra{0}
\end{align}
where for fermions $r=\operatorname{atan}\left(e^{-\pi c\,\omega/a}\right)$, which then allows us to write (as per eq. (\ref{bosonstate})) the state of the Dirac field as
\begin{align}
    \rho_{AI}=\frac{1}{2}(\ket{0}\bra{0}\otimes R_{00}+\ket{1}\bra{1}\otimes R_{11} \nonumber \\
  +  \ket{0}\bra{1}\otimes R_{01}+\ket{1}\bra{0}\otimes R_{10})
\end{align}
This form and the properties of the \(R_\sigma\)'s allows the steps outlined in the bosonic case to carry through; we need only replace the traces of the \(M_\sigma\) matrices with those of the \(R_\sigma\) matrices from which it is straightforward to see
\begin{align}
    D^A_G = \min_{x_3}[\frac{1}{4}((\cos^2 r + \cos^4 r) +x^2_3(\cos^2 r - \cos^4{r}))]
\end{align}
which is clearly minimized at \(x_3=0\) giving
\begin{align}
    D^A_G =\frac{1}{4} (\cos^2 r +\cos^4 r )
\end{align}

Calculation of the discord in the case that Rob does the measurement is once again more difficult than the case where Alice does the measurement because his state is not the maximally mixed state. The probabilities of each measurement outcome are in general functions of \(r\) and \(\mathbf{x}\); this complicates the calculations. However it can be shown that minimization occurs with \(x_3=0\) for which it is simple to verify

\begin{align}
    D^R_G = \frac{1}{4}(1+\cos^2 r(2\cos^2r-1))
\end{align}

\section{On the geometric discord based on different norms and the practical ability to distinguish two states at zero relative distance} \label{appendix}

In this section we analyse the behaviour of other possible geometric measurements of quantum discord for the infinite dimensional states that appear in this paper. We will show that the classical state that we use to obtain an upper bound on geometric discord has a non-zero trace-norm distance. We will then discuss our ability to distinguish the state $\rho_{AR}$ from $\chi$ in the limit of infinite acceleration when their Hilbert-Schmidt distance vanishes.

If we redefine the geometric discord \eqref{defdisc} as
\begin{align}
    _1D^A_G(\rho_{AR}) \equiv \min_{\chi \in \mathcal{C}}||\rho_{AB}-\chi||_1
\end{align}
 with $ ||\rho_{AB}-\chi||_1$ being the trace norm instead of the Hilbert-Schmidt norm
\begin{align}
    ||A||_1 \equiv \text{Tr}(\sqrt{AA^*})
\end{align}
and if we evaluate an upper bound to this new 1-geometric discord using the same prescription that we made to provide an upper bound for the regular discord based in the H-S norm we obtain the following  expression
\begin{align}
  \left.  _1D^A_G(\rho_{AR})\right|_{x_3=1} =\sqrt{(1-t)^3}\sum_{n=0}^\infty t^{2n}\sqrt{n+1}
\end{align}
This can be further simplified to
\begin{align}
  \left.  _1D^A_G(\rho_{AR})\right|_{x_3=1} =t^{-2}\sqrt{(1-t)^3}\text{Li}_{-\frac{1}{2}}\!\left(t^2\right)
\end{align}
where $\text{Li}_{s}\!\left(x\right)$ is the polylogarithm of order $s$
\begin{equation}
\text{Li}_{s}\!\left(x\right)=\sum_{n=1}^\infty\frac{x^n}{n^s}
\end{equation}
which is indeed monotonically decreasing. However  in the infinite acceleration limit takes a finite value $\lim_{a\rightarrow\infty}\left._1D^A_G(\rho_{AR})\right|_{x_3=1} \approx 0.31$.

This does not mean that the 1-geometric discord that we have just defined is not zero in the infinite acceleration limit. However it does
mean that,  at least for the same classical state that provided an upper bound equal to zero for the standard geometric discord in this limit, the 1-discord is non-zero. So are we then permitted to say that this state would not be distinguishable from a classical state in a given experiment?

One can compute the difference between the same observables evaluated in $\rho$ and $\chi$ to try to evaluate if the states that are infinitely close to each other in the H-S norm are possibly distinguishable in an experiment. The naive idea exposed above would be that two states whose distance is zero should be indistinguishable. While this is true in finite dimension, we will show that it is not true in our case.

 It is trivial to see that as $\chi$ is just the diagonal part of $\rho$, all the local observables will be the same for both states. So $\rho$ and $\chi$ are indistinguishable by means of local measurements.

 One could ask what happens with  non-local observables. It is trivial, since $\chi$ is just a diagonal matrix, that the canonical position and momentum correlators are zero, this is
\[\langle X_AX_R\rangle_\chi=\langle P_AP_R\rangle_\chi=0 \]
 where $X=2^{-1/2}(a+a^\dagger)$ and $P= (\text{i}/\sqrt{2})(a^\dagger-a)$.

If we use \eqref{bosonstate} and evaluate $\langle X_AX_R\rangle_\rho=\text{Tr}(\rho_{AB}X_AX_R)$ we obtain that
\begin{align} \label{step1}
  \langle X_AX_R\rangle_\rho= \frac{1-t^2}{2\sqrt{2}}\text{Tr}\left(X_RM_{01}+ X_RM_{10}\right)
\end{align}
Given that (see \eqref{Ms})
\begin{align*}
\text{Tr} X_RM_{01}=\text{Tr} X_RM_{10}=\sqrt{\frac{1-t^2}{2}}\sum_{n=0}^\infty (n+1) \, t^{2n},
\end{align*}
which yields
\begin{align} \label{step3}
    \text{Tr} X_RM_{01}=\text{Tr} X_RM_{10}=\sqrt{\frac{1-t^2}{2}}c^4,
\end{align}
where $c\equiv\cosh r$. Substituting in \eqref{step1}
\begin{align}
  \langle X_AX_R\rangle_\rho= \frac{(1-t^2)^{3/2}}{2}c^4
\end{align}
which is clearly divergent when $a\rightarrow\infty$.

Hence in principle, even when two states are infinitely close to each other in the H-S distance, there exist observables that can be divergently different. In this case, the $X$ correllators are zero for the classical state $\chi$, while they are divergent for the state $\rho_{AR}$ arbitrarily close to it in the H-S norm.

That said, the $X$ correlators are unbounded operators, and it is still not clear how one could effectively measure such correlations in an infinite dimensional system as the one that appears here. Clearly, no apparatus  can in practice distinguish a state of $N$ photons from a state of $N+1$ photons when $N$ is very large; and in the infinite acceleration limit, the relevant Fock components of the state $\rho_{AR}$ are precisely those with very large $N$, hindering the possible practical access to correlations in such a state (indeed, for large accelerations one can check that the correlations in the partial systems of limited number of photons decay to zero).

This indicates that one has to be careful when analyzing infinite dimensional systems of unbounded energy when it comes to the study of quantum correlations. We are definitely safe stating that the quantum tasks using (standard) geometric discord as resource cannot be carried out, even in theory, in the infinite acceleration limit, whereas some other correlations may survive such limits even though its existence cannot be acknowledge via any practical protocol.

\section{Conclusions} \label{conclusions}

In this paper we have analyzed the geometric quantum discord in one of the most common scenarios where entanglement in non-inertial frames has been historically studied. We have shown that in these scenarios peculiar states which present non-zero discord but vanishing geometric discord naturally appear. 

While the entropic notion of quantum discord between bosonic fields remains non-zero
in the infinite acceleration limit, we have shown that the geometric discord actually vanishes, indicating a qualitative distinction between these two
measures. This implies that the behavior of quantum correlations in non-inertial frames (and by extension, gravitational fields)  is somewhat more subtle
than might be originally expected. We have argued that our result has two implications. Firstly, the Alice-Rob scenario acts to illustrate that the geometric discord is not a faithful measure of quantum correlation when considering infinite-dimensional systems with large mean energy. Secondly, the vanishing of geometric discord nevertheless implies a significant limitation on the usable quantum correlations for large accelerations, specifically towards the use of quantum communication protocols such as teleportation and remote state preparation.

It should be noted that scenarios of the type we have considered can be extended to discuss quantum correlations in black hole spacetimes \cite{blackhole}. The near-horizon limit of a large class of horizons from the perspective of a static observer is equivalent to the Rindler horizon perceived by an accelerating observer. Even more interesting is the possibility to export this result to dynamical stellar collapse scenarios \cite{blackhole2}, in view of latter results  on quantum correlations behaviour in the presence of a full dynamical stellar collapse \cite{toappear}. With these prospects in mind, work of this sort may be found to have implications for the problem of black hole information loss.

Finally, we would like to note that the results found here are commensurate with those of Êthe behaviour of vanishing quantum correlations of two (initially entangled) accelerated qubits \cite{Celematsas}. Although this study is specific to the detector model Êemployed, Êit further suggests that the correlations in accelerated quantum systems may not survive the infinite acceleration limit

\section*{Acknowledgements}

The authors would like to greatly thank Gerardo Adesso and Marco Piani for the extremely enlightening discussions about general quantumness. Research at Perimeter Institute is supported by the Government of Canada through Industry Canada and by the Province of Ontario through the Ministry of Research \& Innovation.  This work was supported in part by the Natural Sciences \& Engineering Research Council of Canada.


\end{document}